\documentclass{aastex}
\usepackage{spr-astr-addons}
\usepackage{url}

\RequirePackage{color}

\begin{document}

\title{Line shifts and sub-pc super-massive binary black holes\\ }
\slugcomment{Not to appear in Nonlearned J., 45.}
\shorttitle{Line shifts and SMBBHs}
\shortauthors{Simi\'c \& Popovi\'c}

\author{Sa\v sa Simi\'c\altaffilmark{1}} \and \author{Luka \v C. Popovi\'c\altaffilmark{2}}
\affil{Astronomical Observatory, Volgina 7, Belgrade}

\altaffiltext{1}{Faculty of Science, Radoja Domanovica 12, Kragujevac, Serbia.}
\altaffiltext{2}{Astronomical Observatory, Volgina 7, Belgrade, Serbia.}

\begin{abstract}
Here we discuss the possibility of super-massive binary black hole (SMBBH) detection, using
the shifts of the broad lines emitted from a binary system. We perform a number of simulations of
shapes and shifts of $H_{\beta}$ lines emitted from
SMBBHs, taking into account the emission from two different regions located
around both black holes, and kinematical effects which should be present in a SMBBH. In the model we connect
the parameters of the lines with the mass of black holes and find that the peak shift depends,
not only on kinematical effects of system rotation
and black hole mass ratio, but it is also a function of the individual masses of the two black holes (BHs).
\end{abstract}


\section{Introduction}

It is widely accepted that mergers play an important role in the galaxy formation and evolution, and mergers of
supermassive black holes (SMBHs) are expected to be present at the centers of a number of galaxies \citep[][]{Begelman80}.
In the case that a super-massive binary black hole (SMBBH) system is surrounded by gas, one can expect
that observational effects may be present as a combination of the accretion onto one or both of the
SMBHs and the dynamics of
the binary system \citep[see][]{Popovic12}. In some cases, this could result in a binary Active Galactic
Nucleus (AGN) system, that is similar to a single AGN, i.e. the system emits a broad band electromagnetic spectrum
with  broad and narrow emission lines. Due to the dynamics and composition of the two line emitting regions,
however, the line profiles and shifts are expected to
show some specific characteristics \citep[see][]{Popovic12}.

Based on the analysis of Very Large Telescope integral field spectroscopy and Hubble Space Telescope (HST) imaging
SMBBHs are detected on a kpc scale \citep[see e.g.][]{Woo13,Woo14}, but
there is a problem to detect the sub-pc scale SMBBHs. It seems that sub-pc SMBBHs may be detected only using the
spectral characteristics of a binary system \citep[for a review, see][]{Popovic12,bog15}.

The broad lines emitted from an AGN are used to probe of the geometry of the
broad line region (BLR). The BLR seems to be close to the central SMBH,
since the reverberation technique gives the sizes of the BLR as an order of several light days
to several light weeks \citep[see][]{Kaspi05}).
There is a group of AGNs that emit very broad and complex line
profiles, which could be interpreted as two displaced peaks, one blueshifted and one redshifted from the systemic velocity defined by
the narrow lines, or a single peak shifted line. It has been proposed that such line shapes could indicate a SMBBH
system \citep[see e.g.][etc.]{Gaskell83,Popovic00,Shen10,Tsalmantza11,Eracleous12,Popovic12,Bon12,Liu14,bog15,ru15}.

Here we discuss how the presence of a SMBBH would affect the shift of broad lines and
what might be the observational consequences. First we describe our model of a SMBBH, and after that we analyze a
number of simulations (combining different masses of components and different dynamical effects). The simulations have been
performed to investigate line shift effects in broad lines due to emission from SMBBH. We specifically
incorporate some empirical relationships between the size of the gas, line intensity and BH masses in the model.

\subsection{SMBBH and broad line shift}

The idea that AGN activity is
triggered by mergers is not new \citep[see][]{ko68a,ko68b}, however it is hard to
spatially resolve (i.e. direct imaging) a binary system at sub-pc scale at the center of an AGN. Therefore, spectroscopy
is a tool to study some
effects of a SMBBH in the centers of AGNs.
Spectroscopic studies may provide a method for identifying binary AGNs at sub-pc scale, through the search
of velocity shifts in the complex broad line profile that is
caused by orbiting the two BHs around common center of mass \citep[see e.g.][]{Gaskell83,Boroson09}.
However, it is interesting that of those AGNs, which have been detected as SMBBH candidates by
spectroscopic methods, seem to be controversial because the line shifts and shapes may be explained by other phenomena in the BLR
\citep[as e.g. double-peaked lines with
disc emission, or strong shift and asymmetry due to inflows/outflows][]{Eracleous97,Gaskell10}  and/or there
are some disagreements with other observational facts
\citep[e.g. long-term variability does not show expected the line profile variability][]{Decarli10a,sh15}.

First \cite{Gaskell83} discussed the shifted emission lines in light of the SMBBH hypothesis and reported about
two quasars \citep[see Fig. 1 and 2 in the paper of][]{Gaskell83}. The broad lines in the quasars 0945+076 and 1404+285 were
off-centered (shifted with respect to the systemic velocity) to -2100 km s$^{-1}$ and +2700 km s$^{-1}$, respectively.
The idea about SMBBHs in the center of AGNs has come into focus again nearly three decades later. There is a number
of AGNs that show shifts in their broad emission lines and are good candidates for the
SMBBH scenario \citep[see][]{Tsalmantza11,Eracleous12,ru15}.
In the framework of a SMBBH system there are some expected long term effects that apparently are not
confirmed by observations. Let us mention here two interesting
cases with large shift which may indicates SMBBH system. The first case is the AGN
4C+22.25 where \cite{Decarli10a} found that the H$\beta$ and H$\alpha$
lines show very broad line profiles (FWHM $\sim$ 12,000 km s$^{-1}$), faint fluxes, and extreme offsets
(due to the redshift) (8700$\pm$
1300 km s$^{-1}$) with respect to the narrow emission lines, but the line profiles do not vary in a period of 3.1 years
\citep{Decarli10a}. The second example is the quasar E1821+643 where the broad Balmer lines are
redshifted by $\sim \rm 1000 km s^{-1}$ relative
to the narrow lines and they have high red asymmetry \citep[][]{Robinson10,sh15}. Recent monitoring
of the same source (performed over an interval of 20 years), however, showed
that there is no high variability in the line profile shapes and shifts, which do not
agree with the predicted long-term variability \citep[see][]{sh15}.
In spite of observational facts, one can expect that the line profiles in a long term period should vary \citep[see][]{Bon12,Popovic12}.

We should mention here that the highly shifted broad emission lines may indicate two possible scenarios of a SMBBH:
i) the case where in a SMBBH system only one component has a BLR, and ii) emission of the BLR bound to a recoiling SMBH, i.e., in the stage following coalescence. There are several
AGNs with a relatively large shift of the broad line components \citep[see e.g.][]{Tsalmantza11,Eracleous12,ru15}. Among them
SDSS J092712.65+294344.0 was the first candidate for a recoiling SMBH \citep[][]{Komossa08}, with a broad component blueshift
of 2650 km s$^{-1}$ relative to the narrow emission lines. One can expect that both SMBH have emitting region.
In this case there can be broad lines with specific shapes \citep[see][]{Bon12}, or lines with two peaks. Here we will discuss only the
case where we have a binary system which has one or two broad line emitting regions. We will not consider recoiling black holes.
Here we perform a number of models taking into account the dynamical and physical effects connected with the masses of black holes and their separations.

\section{Model}

In order to explore the shift of the total broad line emission, which is the composite of the emission
from two BLRs located around the SMBH members of a binary system, we propose a relatively simple model,
which takes into account the dynamics of the SMBBH similarly to \cite{Popovic00}  and \cite{Shen10}. Moreover,
our model incorporates the properties of the BLRs around BHs in more detail; the broad emission line parameters
are
correlated with the masses of the central BHs, as revealed by studies
involving reverberation techniques \citep[][]{Wu04,Kaspi05,pet14}.

\subsection{Dynamics of SMBBH}
\label{sec:dynamicsSMBBH}

Basic theory of motion of binary systems is very well established. Here we consider the general case where two components
with masses $m_1$ and $m_2$ and eccentricity $e_1$ and $e_2$ revolve around common the center of mass located at a common focus of the elliptical orbit of BHs. In such case the major axes of the ellipses are:
$$a_1=\frac{R}{(1+q)(1+e_1)}$$
and
$$a_2=\frac{qR}{(1+q)(1+e_2)},$$
where $R$ presents the maximum separation between components and $q$ denotes the component mass ratio
$m_1/m_2$. At any particular moment during the orbital period one can find the position of the components with respect to the barycenter using the basic equation of the ellipse:
$$r_i(t)=\frac{a_i(1-e_i^2)}{1-e_i \cos(\omega t)}.$$

The period of rotation of such system $P_{orb}$ and the orbital velocities of components $v_i(t)$ are \citep[see][]{Paczynski71,Yu01}:
\begin{equation}
P_{orb}=210 \left(\frac{R_S}{0.1 \rm pc}\right)^{3/2} \left(\frac{2\times 10^8M_{\odot}}{m_1+m_2}\right)^{1/2} \rm yr
\label{eq:Porb}
\end{equation}

\begin{equation}
v_{i}(t)=1.5\times10^3 \sqrt{\frac{0.1 \rm pc}{r(t)}}
\sqrt{\frac{m_1+m_2}{2\times 10^8M_{\odot}}} \left[\frac{2 m_1 m_2}{m_i(m_1+m_2)} \right] \cdot\mu_i
\label{eq:vi}
\end{equation}
with $i$ denoting the two components ($i=1,2$), $R_S$ mean component distance, $r_i(t)$ represents the $i-$th component position relative to the common focus as a function of time $t$. The parameter $\mu_i$ replace expression $\mu_i=\sin(\theta)\cos(\omega t+(i-1)\pi)$, where $\theta$ is the angle between the normal to the orbital plane and the line-of-sight of an observer (inclination of the orbit) and $\omega$ angular velocity of the system.

We assume that each component has its own accretion disc that represents a source of emission able to
produce broad lines in the BLR.

\subsection{Physical properties of BLRs and broad line parameters}
\label{sec:kinematicsBLR}

In the standard AGN model a large region of moderate density ($n_e\sim 10^{9} \rm{cm^{-3}}$) surrounds the accretion disc,
also known as the Broad Line Region -- BLR \citep[see e.g.][]{Sulentic00}.
We suppose that most of the BLR is distributed in same plane as accretion disc itself, with inner parts
overlapping the accretion disc and spans a few tens of light days in diameter \citep[][]{Kaspi05}.
The kinematics of the BLR can be very complex, but it is clear that the properties
of the BLR depend on the parameters of the central black hole \citep{pet14}. Here we use some results obtained from reverberation
to constrain the BLR parameters, taking into account the BHs masses in the binary system. We will consider the H$\beta$ line,
since the line has been widely used in black hole determination and in the monitoring campaigns. We also suppose,
that other optical lines, such are $FeII$ or $MgII$, if they are generated by same
BLR cloud, mimic the behaviour of the H$\beta$ line, although similar analyse of their behaviour could give additional
results for our computations. For the reason of simplicity we will suppose the Gaussian line profile instead of Lorentzian,
since we are particulary interested in analyzing the line shifts which does not depends
much on the line profile itself.

We should note here that the BLRs of two interacting BHs (within the small separation $(<1pc)$)
may be significantly affected by the dynamical interaction, and it may cause that the BLR kinematics in
each BLR can be different that one observed in the case of the single central  BH. To avoid it, we take that the BLRs
are significantly smaller than the distance between two BHs, and assume that the gravitation of a black hole
is dominant for the local BLR. In that case we can apply the empirical relationships obtained from the case of
single BH. However, there may be some gas around the SMBBH that can contribute to the line profile, that we
neglected in this model, since only a detailed magnetohydrodynamical study can give the gas distribution around such complex system,
that is out of scope of this work. Moreover, such a component probably will contribute to the center of line (expected
smaller dispersion velocity since the emitting gas in this region is not too close to BHs as the local BLRs), and will
have negligible influence on the total line shift.

\subsubsection{Estimation of the BLR sizes}

We estimate the BLR size by using the empirical formulas \citep{Kaspi05}:
\begin{equation}
\frac{R_{BLR}}{10\textrm{ lt-days}}=(2.21\pm0.21)\left[\frac{\lambda L_{\lambda}(5100{\textrm{\AA}})}{10^{44}\textrm{ ergs s}^-1}
\right]^{0.69\pm0.05},
\label{eq:rblr}
\end{equation}
where $\lambda L_{\lambda}(5100{\textrm{\AA}})$ is the total disc luminosity at 5100$\textrm{\AA}$; for different wavelengths different
BLR dimensions can be computed, e.g., utilizing formulas given by \cite{Kong06}. Here we calculate the continuum emission at
5100 \AA\ for each of the two accretion discs. Given empirical dependence Eq. \ref{eq:rblr}
is derived for the case of the low and intermediate luminosity quasars, and we will restrict
our simulations for this case. Further computations could be conducted in the same way on the high-luminosity/high redshift quasars
using the empirical relations presented in the \cite{Kaspi07}.

The disc has the inner radius  $\rho_{in}\sim$ several $R_g$ and outer radius $\rho_{out}\sim R_{RL}$
($R_g$ is gravitational radius for each black hole separately given by $R_g^i=\frac{Gm_i}{c}$). The outer radius can be estimated
as the radius of the Roche lobe of two bodies \citep[see][]{Eggleton83}
$$R_{RL}(x)=\frac{0.49 R_s x^{2/3}}{0.6 x^{2/3} + Log[1 + x^{1/3}])},$$
where $x$ is defined by the mass ratio of the components as follows: $x=q$ for the more massive BH and $x=1/q$
for the less massive one.
The inner radius of the disc is not exactly equal to the one $R_g$ since
the radiation pressure is huge within one $R_g$, producing impossible conditions for
any material to condensate and undergo laminar motion.
This effect is known as Advection-Dominated Accretion Flow (ADAF) and it is present in the case of
the spinning black holes described with the Kerr metric \citep[see][]{Narayan94,Manmoto10}.
For this reason we adopted that $\rho_{in}^i=3.5 R_g^i$ as inner radius of the disc. On the other hand
the outer disc radius is just a half of the corresponding Roche radius since the disc
can not entirely fill the space defined by $R_{RL}$ \citep[see][]{Yan14}.

The distribution of temperature over the radius of the disc follows the power law which in simplified form is given as
$T_{eff}(\rho)\sim \rho^{-3/4}$ \citep{Shakura73}. The emission model is assumed to be thermal in nature
and the amount of radiation generated by other mechanisms (inverse Compton or synchrotron) is thought to be negligible
for optical and UV bands. In order to compute the total amount of radiation coming from the
surface of the disc one needs to integrate emission of the entire disc area. This allows the computation of the specific luminosity $L_{\lambda}(5100{\textrm{\AA}})$ that is used in Eq. \ref{eq:rblr}.
Consequently, we connect the size of the BLR with the individual masses of the two BHs and also with their mass ratio.

\subsubsection{The intensity of the H$\beta$ line}

The H$\beta$ line intensity is estimated using empirical relation given by \cite{Wu04}:
\begin{equation}
\log R_{BLR}(\textrm{lt-days})=1.381+0.684 \cdot
\log\left(\frac{\lambda L(H\beta)}{10^{42}\textrm{ergs s}^{-1}}\right).
\label{eq:lumHb}
\end{equation}
We define $\lambda L(H_{\beta}) \equiv I_{\lambda 0}^{H\beta}$, the luminosity of the $H_{\beta}$ line, and substitute the $R_{BLR}$ with its
coresponding expression from Eq. \ref{eq:lumHb}. This allow us to identify the individual contributions of the two
components to the total line emission in the binary system.

\subsubsection{The width and shift of H$\beta$}

The H${\beta}$ line width is supposed to be
proportional to the black hole mass, since more massive BHs induce higher velocities in the BLR  \citep[see][]{pet14}.

The velocity distribution in the BLR  can be estimated based on the viral theorem as:
$$v_{BLR}(m_i)=\sqrt{\frac{Gm_i}{R_{BLR}}},$$
giving the spectral line width dispersion as
$\sigma_i=\lambda_{H\beta}\frac{v_{BLR}(m_i)}{c}$. Note here that the velocity of gas in the BLR is a direct function of
the associated BH mass and it is not directly tied into the mass ratio $q$. This will be
very important in computing line shifts, since different mass combinations can produce the same mass
ratio, but different orbital velocities of binary components, consequently
different line shifts and velocity dispersion.

The line shift caused by gravitational effects in the close vicinity of the BH can be computed by
$\Delta\lambda_g^i=\frac{Gm_i}{cR_{BLR}}$. For the BLR sizes adopted in our computation, however, the produced line-shift $\Delta\lambda_g^i$ is small compared to
the shifts produced by the radial velocities of binary components.
Consequently, one can neglect the effect of gravitational line shift from the line shape computation.
On the other hand, one should include the shift of lines emitted from different components taking into account
the dynamics of the system. We take that rest wavelength of lines is shifted according to
$$ \lambda = \lambda_0(1+z_{dopp}^i),$$ where $z_{dopp}^i$ represents the ratio $v_r^i/c$, particulary for each component of the binary system.

\subsubsection{The line shapes}

We calculate line shapes of different configurations of SMBBH system as:

$$I_{tot}(\lambda)=I_1(\lambda)+I_2(\lambda)$$
where we assume that each component emits Gaussian line profile:

\begin{equation}
I_{i}(\lambda)=I_{i}(\lambda_0)exp{\left[-\left(\frac{\lambda-\lambda_{0}\cdot (1+z^{i}_{dopp})}{\sqrt{2}\sigma_{i}} \right)^2 \right]}\cos(\theta)
\label{eq:line}
\end{equation}
with $i$ designating the components 1 and 2, $\lambda_0$ wavelength of H${\beta}$ line in laboratory frame, and $z^i_{dopp}$ Doppler corection for radial component velocities.
We also ascribe a flat distribution to the BLR, coplanar with the orbital planes of the SMBBHs and therefore defined by the same inclination angle $\theta$.

\subsubsection{Model constraints and measured line parameters}

First of all, using this model, one has to adopt some constraints: a) since we derive the radius of the BLR from the
continuum luminosity, one should check if the size of the BLR fit within the separation between the two BH components. In order to avoid semi-detached
system we constrain our model that $R_{BLR1}+R_{BLR2}<R$. In first approximation, even if there are small interactions of BLRs, it should not had much influence on the dynamics of binary system, as well as on the emitted amount of radiation, since the interacting volumes are expected to be quite small in comparison with the whole region of BLRs.
After several tests, we obtain that the limiting value of $R$ in considered models is 0.05pc for masses of the order of $10^8M_{\odot}$. b) As we investigate the role of BH mass on the line shapes and
shifts, we explore the range of masses $10^8 - 10^9\ M_\odot$ for the two BHs. c) The orbital parameters as well as the mass ratio affect the total line shift \citep[see][]{Popovic00,Shen10}. To explore individual mass effect we consider the cases q=0.25, 0.5 and 1, and we assume eccentricities $e_{1,2}=e_1=e_2=0.7$.

We would like to get time-scale estimates for an orbital period, which can be used to predict the
long-term effects on line-shifts to compare against long-term monitoring campaigns for SMBBH candidates.
Thus, we calculate the time for different phases (specific snapshots) and the corresponding separations between BHs,
testing various mass ratios and maximum separations between the components.

As one can see from Table \ref{tbl:orbit_pars}, the changes in the line shift (caused by dynamical effects)
can be present in a time interval of the order of decades. The calculation for periods and distances has been performed for a SMBBH system
with mass of the smaller component equal to $10^9M_\odot$ and a distance between components of $R=0.1$ pc.

\begin{table}[ht]
\begin{center}
\caption{Computed values for duration of orbital phases $t[years]$  and corresponding component separations
$r_{sepp}$ for different dynamical parameters $q$. The  eccentricity of the orbits is taken as $e_{1,2}=e_1=e_2=0.7$.}
\vskip 4mm
\begin{tabular}{|c|c|c|c|c|}
\hline
\emph{parameters}& $P_{orb}/4$ & $P_{orb}/2$ & $3P_{orb}/4$ & $P_{orb}$ \\
\hline
q = 1 & 14.2 & 28.6 & 42.8 & 57.1\\
q = 0.5 & 11.6 & 23.3 & 35.1 & 46.6\\
q = 0.25 & 9.1 & 18.0 & 27.1 & 36.1\\
\hline
\hline
$r_{sepp}\rm (pc)$ & 0.03 & 0.017 & 0.03 & 0.1 \\
\hline
\end{tabular}
\label{tbl:orbit_pars}
\end{center}
\end{table}

\begin{figure}[h]
\plotone{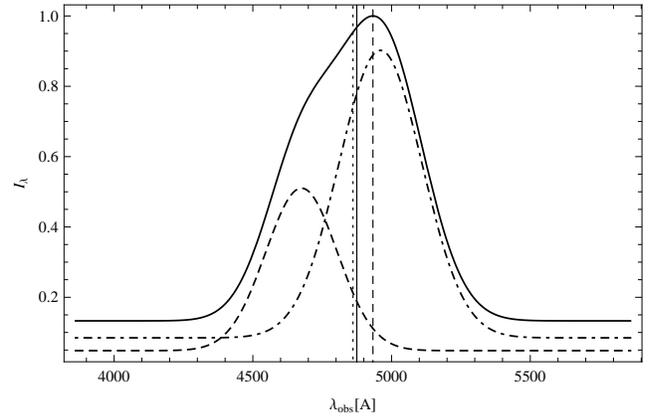}
\caption{H${\beta}$ line for a BBH system, with mass ratio $q=0.5$ and particular component masses of $m_1=4\cdot10^8$ and $m_2=8\cdot10^8 M_\odot$.
Solid curved line shows a total H${\beta}$ line from component 1 (dashed curved line) and component 2 (dot-dashed curved line).
Vertical straight lines designate unperturbed wavelength (dotted line), shift of the peak (dashed line)
and centroid (solid line) of combined H${\beta}$ line. Note: Centroid shift is intentionally slightly displaced from it's computed position for the reason of clarity.}
\label{fig:Hb}
\end{figure}

To explore the shift in SMBBHs and its usage for detection of binary black holes we measure the shift of the peak (the maximum of the line).
Additionally, we compute the wavelength of the line centroid and appropriate shift as:

$$\lambda_c={{\int{I_{tot}(\lambda)\lambda d\lambda}}\over{\int{I_{tot}(\lambda) d\lambda}}} $$
$$\delta\lambda_c=\lambda_0-\lambda_c $$

The total line shape of the H${\beta}$ line generated by the sub-pc binary system of SMBHs (with masses of $m_1=4\cdot10^8$ and $m_2=8\cdot10^8 M_\odot$)
and parameters $q=0.5$ and $R=0.05\rm$ pc is presented in Fig. \ref{fig:Hb}.
As we can see in this example, the line shows a complex profile with a high blue asymmetry, showing a blue shoulder and a redshifted
peak. We present the line peak and shift of the centroid with vertical lines. In this particular case
the shift of the line peak is significantly larger than the shift of the line centroid.

\section{Results and discussion}
\label{sec:results}

In previous sections we assume that both components of binary system retain their accretion discs and surrounding BLRs.
However, it is of particular interest to investigate a system where just the smaller component has BLR,
since in that case there is no line combination with the other component, and the resulting shift would be more clearly defined.
For that reason we will discuss two cases in this section: \emph{i}) both components have a BLR and \emph{ii}) only the smaller component has a BLR.

\subsection{The case of both black holes with BLRs}
\label{sec:both_disc_BLRs}

\begin{figure*}
\epsscale{2}
\plotone{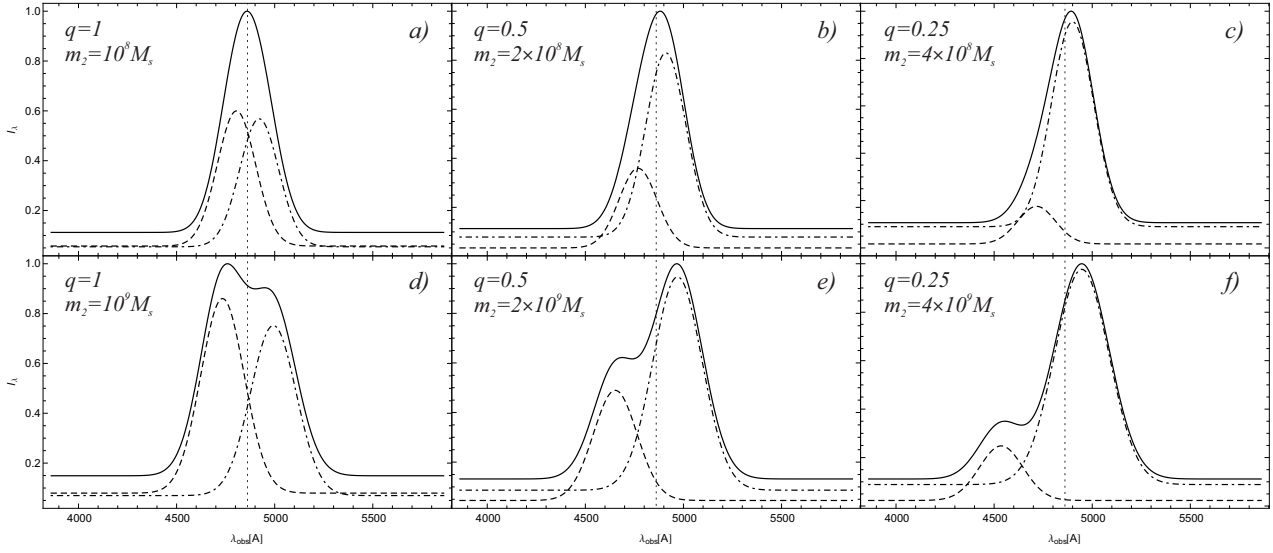}
\caption{H${\beta}$ line shifts for different mass ratio $q$ and mass of components (given on plots).
In first row (panels \emph{a),b),c)}) the maximum BH component separation is
$R=0.05\rm pc$, while in the second raw (panels \emph{d),e)} and \emph{f)}) the maximum separation is $R=0.1\rm pc$.}
\label{fig:R005e07ph4q}
\end{figure*}

In the case of bound BHs orbiting around the common barycenter, their masses determine the dynamics of orbital motion
and consequently the shifts of the components that contributing to the shift of the observed (total) line profile.
As we noted above, we have compute the shape of the H${\beta}$ line for different combinations of masses and mass ratios $q$.
We first consider the case of $q=1,0.5,0.25$, with associated individual BH masses of $m_1=1\cdot10^8M_{\odot}$ and $m_2=1,2,4\cdot10^8M_{\odot}$
(see Fig. \ref{fig:R005e07ph4q} panels \emph{a), b), c)}). We than change the individual BH masses to $m_1=1\cdot10^9M_{\odot}$ and $m_2=1,2,4\cdot10^9M_{\odot}$
while keeping the same $q$ ratios as before (see Fig. \ref{fig:R005e07ph4q} panels \emph{d), e), f)}).
In both cases computations here refers to the quite specific configuration with longitudinal alignment between line-of-sight (LOS) and the axes of the elliptical paths of two BHs, for the reason that in such case
radial velocities over the LOS are highest for the orbital phase $P_{orb}/2$ and given inclination angle $\theta$, with the components at closest separation
(see Table \ref{tbl:orbit_pars}). Consequently a larger line shift during the orbit is expected.

As one can see in Fig.\ref{fig:R005e07ph4q}, when increasing the mass of the second component $m_2$, both the intensity and the width of the associated line go up.
Although the mass ratio is the same (panels \emph{a-d, b-c,c-f}), both the shift and asymmetry of the total/composite line become more prominent in the more massive binary system
(panels \emph{d-e-f} compared to \emph{a-b-c}).
It shows, that beside the dynamical parameters (the mass ratio, phase and separations),
the masses of components have an important influence on the line profiles and determine the detectability of the effect in SMBBH system.
In Table \ref{tbl:shifts_R005e07ph4q} we give measured values of the line shifts for
components (Columns 2 and 3), the total shifts (Column 4) and centroid shift (Column 5). As it can be seen, with
bigger mass of the components the shift stays larger, and consequently the effect of a binary system is more pronounced.

\begin{table*}[ht]
\begin{center}
\caption{Numerical values of shifts presented in Fig. \ref{fig:R005e07ph4q}. First Column denotes the corresponding panel in
Fig. \ref{fig:R005e07ph4q}, second and third Columns the shift of peaks for components, respectively. In fourth and fifth
columns are given shifts of peak and centroid of the total line profile, respectively.}
\vskip 4mm
\begin{tabular}{|c|c|c|c|c|}
\hline
panel & $\delta\lambda_m^1 \rm [km/s]$ & $\delta\lambda_m^2 \rm [km/s]$ & $\delta\lambda_m \rm [km/s]$ & $\delta\lambda_c \rm [km/s]$ \\
\hline
\emph{a)} & -3450 & 3696 & -110 & 32 \\
\emph{b)} & -5711 & 3050 & 1266 & 278 \\
\emph{c)} & -8915 & 2405 & 1935 & 451 \\
\emph{d)} & -7828 & 8156 & -6268 & -110 \\
\emph{e)} & -12881 & 6718 & 6482 & 333 \\
\emph{f)} & -20042 & 5294 & 5282 & 630 \\
\hline
\end{tabular}
\label{tbl:shifts_R005e07ph4q}
\end{center}
\end{table*}

\begin{figure*}
\epsscale{2}
\plotone{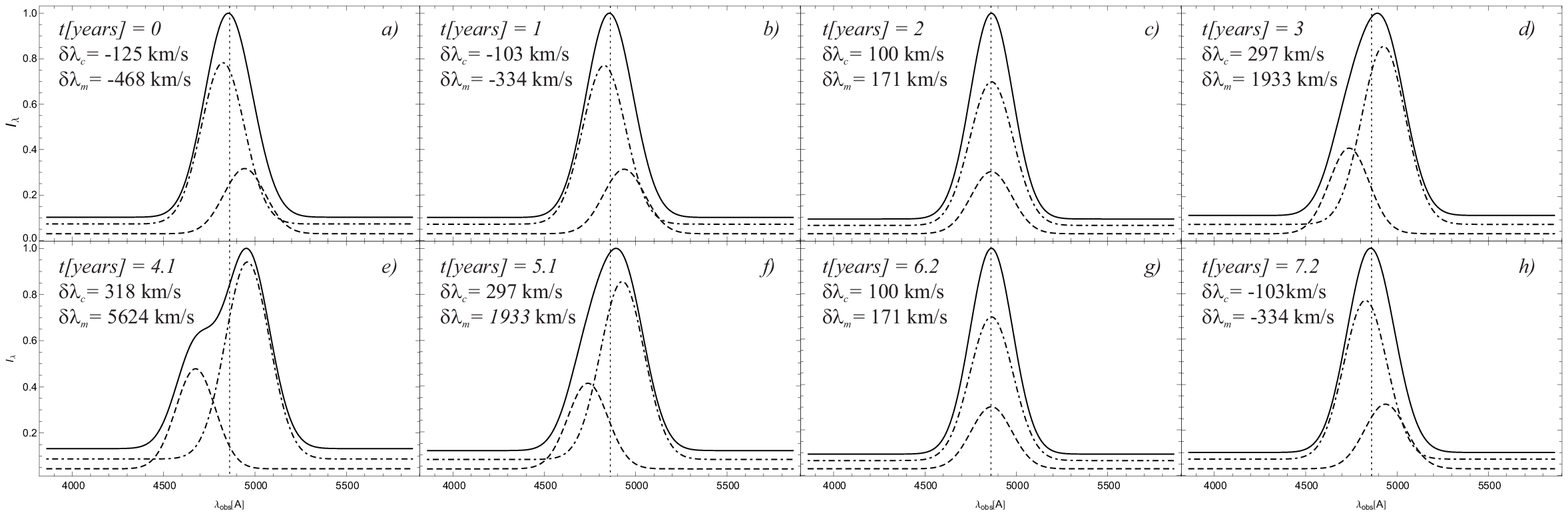}
\plotone{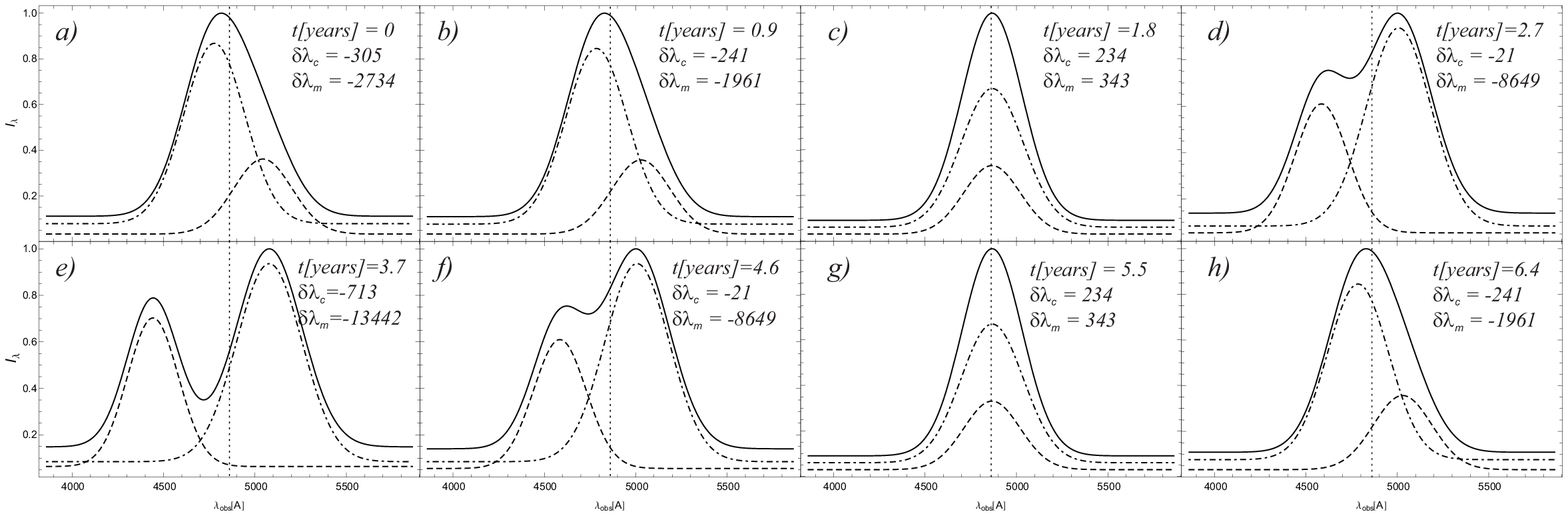}
\caption{The total H${\beta}$ line profiles during full orbits for two SMBBH systems.  Upper panels (from \emph{a}) to \emph{h})
are given for the mass $m_1 = 4\cdot10^8\rm M_{\odot}$ and $R=0.05\rm pc$, and lower panels (from \emph{a}) to \emph{h}) for
$m_1 = 4\cdot10^9\rm M_{\odot}$ and $R=0.1\rm pc$. For both systems we fixed $q=0.5$ and $e_1=e_2=0.7$.
Dashed and dot-dashed lines represents contributions of the components to the total H${\beta}$ lines (solid lines). In
the panels from \emph{a}) to \emph{h}) we indicate the corresponding phase $t$ in years, line centroid shift $\delta\lambda_c$ and
peak shift $\delta\lambda_m$ in km/s on each plot.}
\label{fig:R005e07q05phx}
\end{figure*}

With increasing the orbital eccentricity of BHs line shifts could be more defined in some particular
phases of the orbit with minimum separation between components. For that reason we compute the shape of the total
H${\beta}$ spectral line for different phases, see Fig. \ref{fig:R005e07q05phx}.
In upper/top panels we present the case of low mass components
with masses $m_1=4\cdot10^8M_\odot$, $m_2=8\cdot 10^8M_\odot$ and $R = 0.05\rm pc$.
The lower/bottom panels show the larger mass case with $m_1=4\cdot10^9M_{\odot}$, $m_2=8\cdot10^9M_{\odot}$ and
$R = 0.1\rm pc$.
In both considered cases with increasing the masses of the BHs, the BLR diameter also increase,
so we intentionally increase separation $R$ to avoid mutual interaction of BLRs.
As it is well known \citep[see][]{Popovic00,Shen10} during an orbital cycle, the broad
line profile been changes, the asymmetry and shift
acquiring different values. However, the asymmetry and line profile changes are
expected to be more important in the case where the components have bigger masses ($\sim$10$^9M_\odot$). The most
interesting is that only in one case (panel \emph{e}) for higher mass case) there is clearly seen a double-peak profile
that {\emph may indicate the emission of two BLRs. These simulations suggest that the asymmetric profiles are far
more prevalent then the extreme double-peak ones.

Note here that different geometries could be applied in order to explain the observed broad spectral lines components.
Double peaked lines could be a result of binary system dynamics, but it seems that double-peaked line shapes
could be achieved when a single AGN has two or more broad line emitting regions,
and if the emission of an accretion disc is dominant  \cite[][]{Eracleous97,Popovic04,Bon09,Popovic11}.  In the case that
a SMBBH system is emitting double-peaked line profiles, the distances between peaks has to be changed, and also, single peaked asymmetric
profile can be observed in a monitoring campaigns, while in the case of double-peaked profiles emitted by the disk
one cannot expect a big changes in the shift of the peaks \citep{Eracleous97}.
Additionally, double peaked profiles originated by the disk-like region usually have a boost in the blue peak, and extended
red wing \citep{Eracleous97,Popovic11}, that is not a rule in the profiles of SMBBHs. The most important difference between
double peaked lines from the disk and SMBBHs is  that the line profiles of
SMBBHs are changing dramatically during time (from single to double peaked), that
is not case for the lines emitted from the disk \citep[see the case 3C390.3 in][]{Popovic11}.

\begin{figure}
\epsscale{1}
\plotone{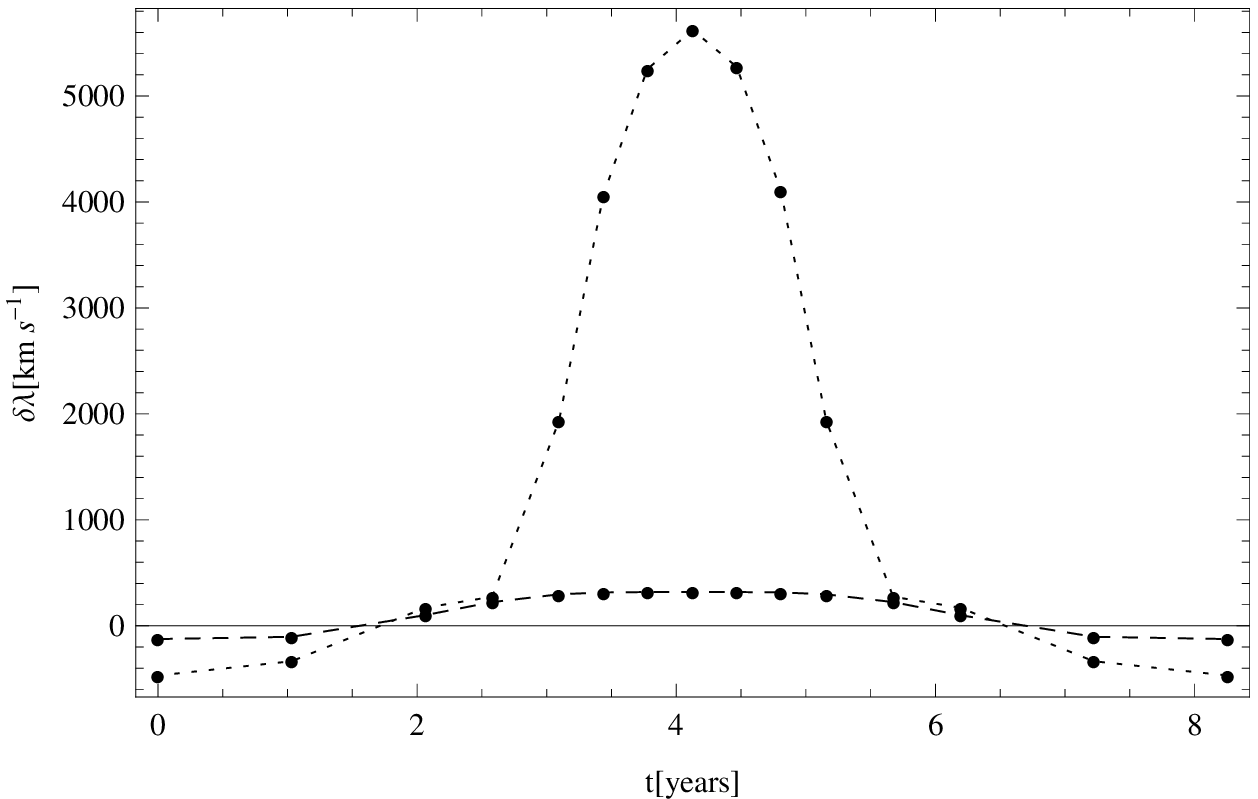}
\plotone{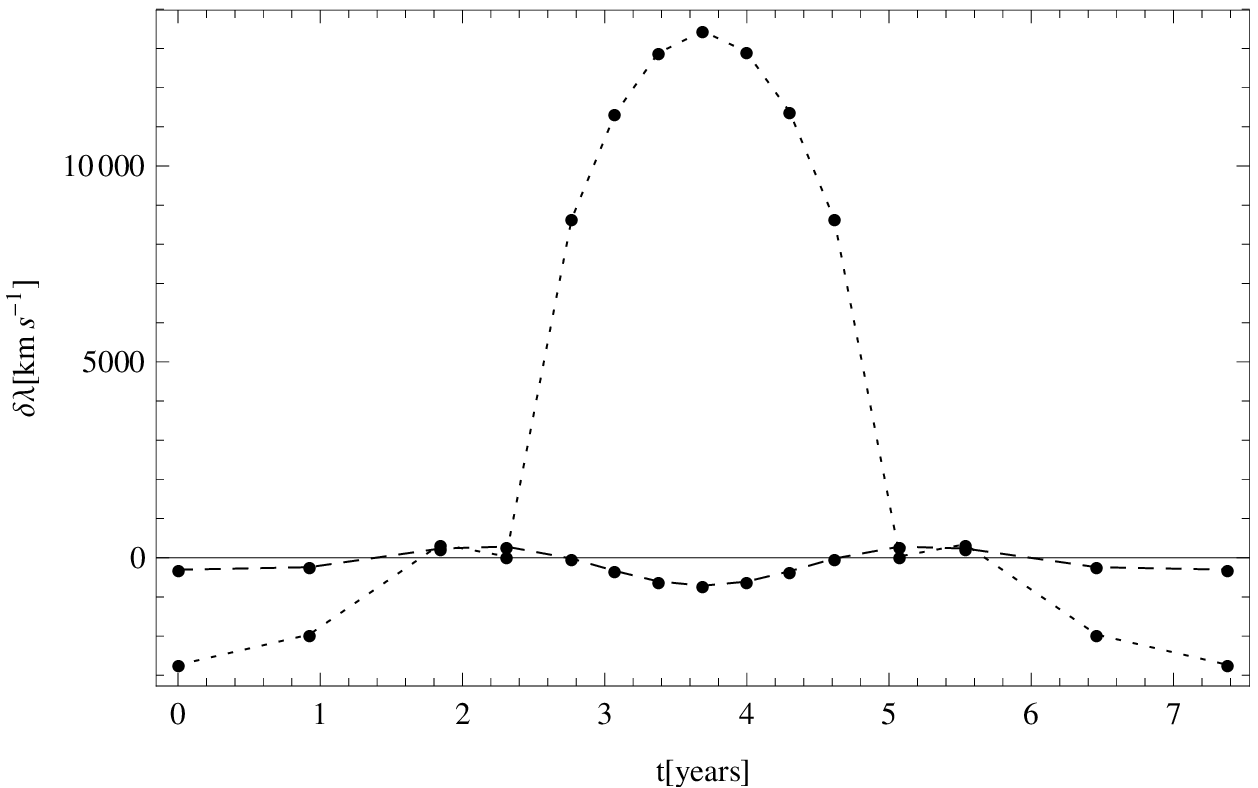}
\caption{The corresponding shifts (doted line) and line centroid shifts (dashed line)
for the cases given in Fig. \ref{fig:R005e07q05phx}.
Upper panel corresponds to the case of lower masses of components (upper panels in Fig. \ref{fig:R005e07q05phx}).}
\label{fig-sh}
\end{figure}

{\emph We should note here that observed broad line profiles in AGN can be complex and specific, as we mentioned above,
the double-peaked profiles can be observed in a group of AGN \citep[around 10\% of AGN, see][]{Strateva03,eh94},
which seems to be emitted from a disc-like geometry. On the other side, it seems that we can divided the AGN using
the width of their profiles
\citep[as e.q. FWHM$>$4000 km s$^{-1}$, see][]{Sulentic09,Zamfir10,Marziani10}
where broader lines show a red/blue asymmetry.
Comparing our, modeled broad line profiles with observations we found that
(see Fig. \ref{fig:R005e07q05phx}) all type of line profiles in the case of a SMBBH is expected.
If there is a SMBBH with smaller masses, we can expect relatively narrow and symmetric broad line
profiles during 2/3 of the orbiting period, in the case of a massive SMBBH, the line profiles are mostly asymmetric.
Therefore, to find more evidence about the structure of emitting system (SMBBH vs. complex BLR), a long-term monitoring,
and analysis of the line profile (shift, width and intensity) variation is needed.

The time evolution of the total line peak shift and line centroid shift for the two cases shown in Fig.
\ref{fig:R005e07q05phx} are presented
in Fig. \ref{fig-sh}. It is obvious that the peak shift is more prominent than the centroid shift. However it is interesting that
only in a relatively short time-window (several years, see Fig. \ref{fig-sh}), the line shift is
significant and could be easily measured, whereas,
for about 3/4 of a period the shift has small values.

{\emph We should note here, that some
previous studies  of a large number broad line AGN seem to suggest that blue-shifted peaks are more
prevalent than redshifted ones \citep[see e.g.][]{Zamfir10,Marziani13} that cannot be explained by SMBBH hypothesis.
This also  supports the fact that the long term observations, which can indicate  the changes in the line shift, may
be used for detection of a SMBBH system.

\subsection{SMBBHs with one BLR}
\label{sec:single_BLRs}

For a compact binary system there is a possibility that just one BH has the BLR
\citep[see e.g.][]{Armitage02,Hayasaki07,Cuadra09,Lodato09,Popovic12}.

Good examples of subparsec BBH quasar are PG 1302-102 reported by \cite{Graham15} and Mrk 231, see \cite{Yan15}, recently suggested via either periodical variations or peculiar continuum
spectrum. PG 1302-102 is described as a binary SMBH pair with a total virial mass of $10^{8.5} M_{\odot}$, with observed period limiting the separation between the pair to 0.01 pc. This means that the system has evolved well into the final parsec scale, with two component significantly different in their mass, as it can be concluded from fitting the spectral lines of Hydrogen series (see their Fig. 4).
Also, Mrk 231 binary is high mass ratio system ($q\sim50$), with unique features in the optical-to-UV spectrum and the intrinsic X-ray weakness.
In both cases the emission from the small BH component dominates and the disk associated with the big BH component emits little ionizing photons.

In such case the observed H${\beta}$ line is a singlet, which could be shifted around
the rest wavelength. It is of particular interest to analyze the case when a component with bigger mass is without BLR. This is more probable scenario
since there is a bigger chance that smaller mass component retain it's accretion disc and BLR.

\begin{figure}[h]
\epsscale{1}
\plotone{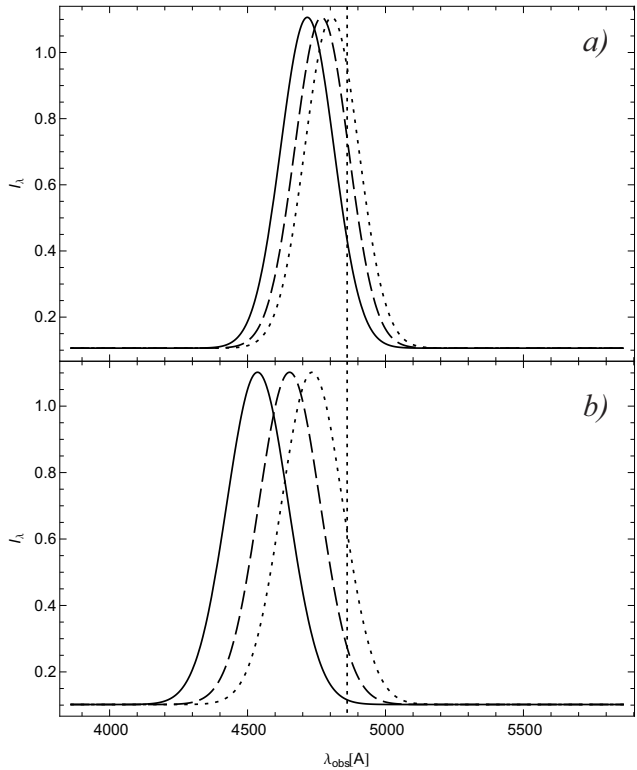}
\caption{The H${\beta}$ line from a single BLR associated with the lower mass BH component of binary system.
Panel \emph{a}) presents the computation for parameter $R=0.05\rm pc$ and the mass of the bigger
component having values $m_2=1,2,4 \cdot10^8M_{\odot}$. Panel \emph{b}) shows the case of $R=0.1\rm pc$ and $m_2=1,2,4 \cdot10^9M_{\odot}$.
Used line designations are: dotted line $q=1$, dashed line $q=0.5$, and solid line $q=0.25$.}
\label{fig:R0105ph4qx1}
\end{figure}

\begin{table}[ht]
\begin{center}
\caption{Centroid shift for single H${\beta}$ line presented in Fig. \ref{fig:R0105ph4qx1}. The first three rows of
data are for the case when $m_2=1,2,4\cdot10^8M_{\odot}$, while last three rows are for $m_2=1,2,4\cdot10^9M_{\odot}.$}
\vskip 4mm
\begin{tabular}{|c|c|}
\hline
$q$ & $\delta\lambda_1^c \rm [km/s]$  \\
\hline
1 & -1844 \\
0.5 & -3047 \\
0.25 & -4745\\
\hline
1 & -4517 \\
0.5 & -7440 \\
0.25 & -11578 \\
\hline
\end{tabular}
\label{tbl:shifts_R0105ph4qx1}
\end{center}
\end{table}

As we can see in Fig. \ref{fig:R0105ph4qx1} and Table \ref{tbl:shifts_R0105ph4qx1} where we present the computations for the case of just a lower mass BH containing it's BLR, the total
shift of the H${\beta}$ line is significantly greater than the shift in the case of the combined line produced by two BLRs.

Also, in this case, we obtain that shifts in the single component depends on the mass, as expected.
However, if we have the same mass ratios, one can expect that the shift will be larger in the higher mass system.
As one can see in Table \ref{tbl:shifts_R0105ph4qx1},
when the mass is increased by an order of magnitude, the shift increases 2-3 times.

Line shifts in the single BLR case is caused by
 the orbital motion of the smaller component.
Our computation in Fig. 5 is performed for the highest radial velocity.
As we can see in Fig. 5, and Table 3,  for both considered  cases
(a) and b)), the shift is, approximately, a linear function of the mass, i.e.
 $\delta\lambda_1^c\sim m_2$.  This
is in accordance with expectations since with
increase of $m_2$, the center of mass of the system is closer to bigger component,
and smaller component will have higher radial velocity, producing higher line shifts.

\begin{figure*}
\epsscale{2}
\plotone{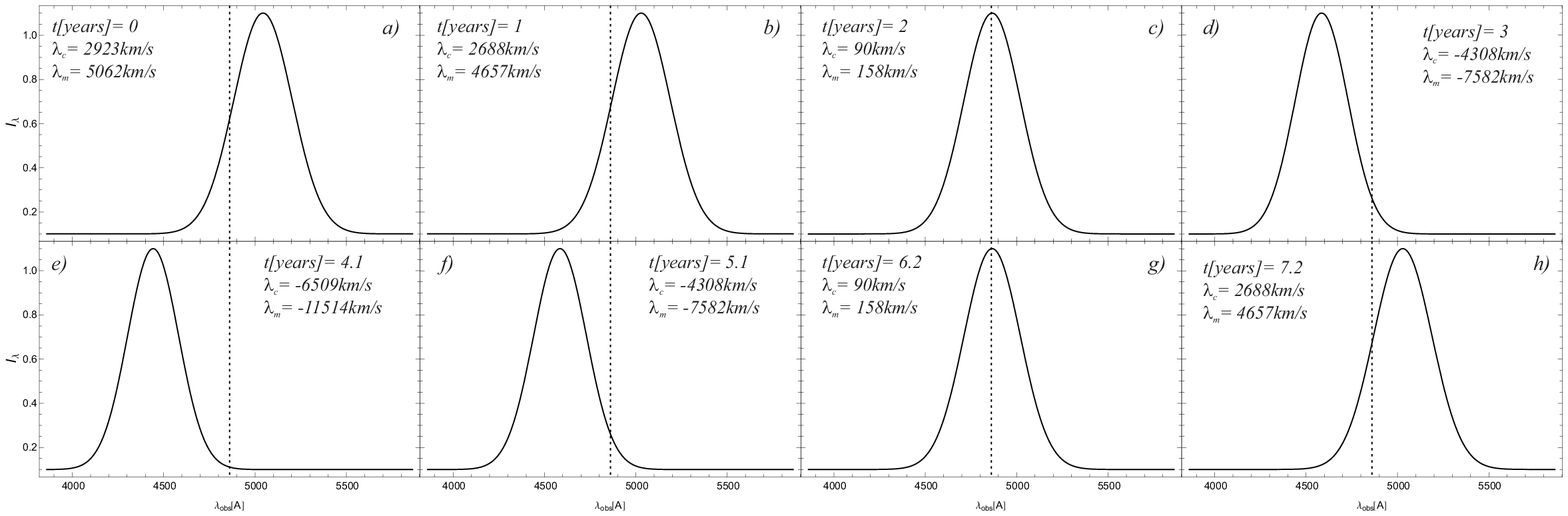}
\caption{The H${\beta}$ line for the case where only the lower mass BH in the binary system retains its emmiting BLR. Given parameter
are same as in Fig. \ref{fig:R005e07q05phx}, but for single line.}
\label{fig:R005e07q05phx_1}
\end{figure*}

In Fig. \ref{fig:R005e07q05phx_1} we show the time-evolution for the line centroid and shift over a full orbital period of the rather peculiar system we simulate.
We note that the line is very symmetric at all phases, so the line asymmetry as shown in the case of two emitting BLRs, cannot be seen.
A system like this cannot account for highly asymmetric lines that show complex profiles.

\section{Conclusion}
\label{sec:concl}

In this paper we consider the possibility to detect the line shift of SMBBH systems. To investigate the shift effect we develop a model,
which is taking into account the dynamical effects similar to \cite{Popovic00} and \cite{Shen10},
but in contrast with models
given in these papers, we include physical properties of the emitting regions in both components of a SMBBH system. We constrain the
radius of the BLR taking into account the emission of the continuum at 5100 \AA\ (estimated from disc emission around components), the
intensity of the H$\beta$ line of each component considering empirical relationship between the H$\beta$
line intensity and the BLR radius,
and finally the line velocity dispersion using the virial theorem. This model allows to explore
time variability of the line parameters (shift, width and intensity) of a SMBBH system during an orbit, and can be applied for
modeling of long term spectroscopic observations of the SMBBH candidates.
We perform a number of simulations, changing the masses of components
exploring how it's (not only the mass ratio that is usually is considered) affect the line profiles. Here we
considered two cases: A) both BHs have a BLR, and B) only the smaller BH has a BLR
From our investigation we can outline following conclusions:

In the case A)

(i) Beside the dynamical parameters and the mass ratio in a SMBBH system, the mass of components has an important role
in the observed line profiles. It seems that low mass components (with total mass smaller than $10^8M_\odot$) will only moderately affect the broad
line profiles. Consequently, the dynamic effect of binary system can be dominated by kinematical effects in BLRs and probably they are not
good candidates for detection of SMBBHs using broad lines. The SMBBHs with bigger masses ($>10^9M_\odot$) seems very promising
for detection using SMBBHs broad lines. The line profiles show a complex shape, and line peak can be significantly shifted.

(ii) The  line shapes are changing during an orbit of SMBBHs showing the double-peaked profile (two separated peaks)
only in a short period in the case of massive black holes. This indicates that the double-peaked profiles may be not characteristic
profiles observed from SMBBH, i.e. one can expect to observe more frequently  the asymmetric profiles from SMBBHs,
and also some shoulders in the line profiles.

In the case B)

(iii) As it is well known, the emission of a single BLR in SMBBHs gives single shifted line, however, the shift and width of
lines are larger in the case of more massive SMBBH components ($>10^9M_\odot$) and they are easier to detect. During a period
of revolution of binaries, one can expect that shift is changing form blue to red or vice versa, therefore the
systematic shift to the blue or red in a period can indicate presence of a SMBBH.

Based on the dynamical time-scale of binary BH system we could say that the probability of detection of such an objects is rather moderate, since the the periods of the orbiting are in range of
few years to few tens of the years. However, other parameters also has big influence, such are individual masses, mass ratio, inclinations, etc., which consequently lower the probability of detection binary system.

As a summary we can conclude that, beside dynamical effects, the total mass of a SMBBH has
influence on the line shift and profile shape. Consequently, the observational effects in the broad lines seem to be present in
massive systems ($>10^9M_\odot$). However it is a question if the effects of a close binary system
can be seen in low mass
BHs, i.e. a large number of SMBBHs may give ordinary broad line profiles which are typical for a single AGN.
Therefore, spectral monitoring of SMBBH candidates can be very useful for confirmation of their binary nature, i.e. the
variation in the broad line profiles (shift, width and intensity) should be present, and this variation has to have a
periodical or quasi-periodical nature
following  the orbiting period of a SMBBH.

\acknowledgments

This work is a part of the project (176001) "Astrophysical Spectroscopy of Extragalactic Objects,"
supported by the Ministry of Science and Technological Development of Serbia.

\end{document}